\documentstyle[11pt,paspconf,epsf]{article}

\begin{document}

\title{Radio observations of gamma-ray blazars}

\author{M. Pohl, A. M\"ucke}
\affil{MPE, Postfach 1603, 85740 Garching, Germany}

\author{W. Reich, P. Reich, R. Schlickeiser}
\affil{MPIfR, Postfach 2024, 53010 Bonn, Germany}

\author{H. Ungerechts}
\affil{IRAM, Avenida Divina Pastora 7, 18012 Granada, Spain}




\begin{abstract}
Starting in 1991 we have performed a regular monitoring 
of EGRET sources with the Effelsberg 100-m telescope and
the IRAM 30-m at Pico Veleta. In comparison with data on a sample of
flat-spectrum quasars which have not been seen by EGRET
we search for correlations of any kind in the behaviour of
FSRQ's in the radio and gamma-ray regime. While there is no
radio-to-gamma-ray luminosity correlation for these sources,
it seems that gamma-ray high states coincide with increased activity
in the radio regime with a strong tendency that gamma-ray outbursts
precede radio outbursts. The gamma-ray spectra seem to harden with
increasing flux level.
\end{abstract}




\section{An example of an individual source: 0528+134}
We will first discuss PKS 0528+134,
the most luminous $\gamma$-ray source known besides $\gamma$-ray 
bursts.
0528+134 exhibits superluminal motion with $\beta_{app}\simeq 4.4$ and 
indications for even higher values (Pohl et al. 1995). Superluminal motion
was expected since given the variability time scales of a day, the
redshift of z=2.07 and the $\gamma$-ray flux in the OSSE range, strong
Doppler boosting is required to satisfy the compactness limit and the
Elliot-Shapiro relation (McNaron-Brown et al. 1995). It has been noted
that the expulsion of new VLBI components may coincide with $\gamma$-ray
outbursts (Pohl et al. 1996).
\begin{figure}
\noindent
\plotone{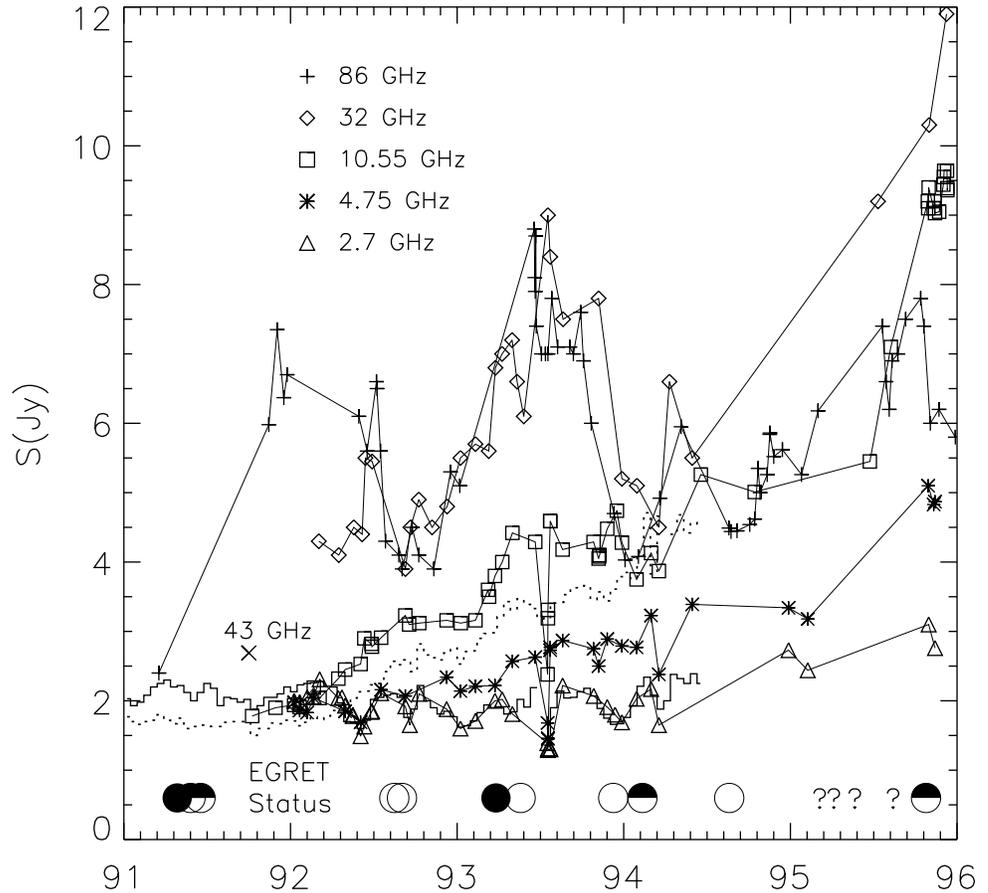}
\caption{Here we show the radio light curve of 0528+134 based on Effelsberg
data and observations at 86\,GHz taken with the IRAM 30-m. The histograms
give the two-weeks-averages of the NRL-GBI data at 2.25 GHz
(solid line) and 8.3 GHz (dotted line). Observational uncertainties
are below 5\% except for the data at 32 GHz and 86 GHz which have
uncertainty levels around 10\%. The unusual depression
in July 1993 is most likely an extreme scattering event and not
intrinsic to the source. For comparison the
state in the EGRET range is indicated by empty circles for low state
($S< 5\cdot 10^{-7}\ {\rm ph.\,cm^{-2}\,sec^{-1}}$ above 100 MeV),
medium level ($S< 10^{-6}$), and
high state ($S> 10^{-6}$). A question mark indicates protected data.
It is tempting to relate the $\gamma$-ray outbursts in 1991 and 1993 to the 
mm outbursts a few month later. However, no strong $\gamma$-ray outburst
has been reported yet for 1995, a few months before the brightest
ever-recorded mm outburst. Or is there a time lag of two-and-a-half
years, relating the 1991 $\gamma$-ray flare to the 1993 radio outburst,
respectively the 1993 $\gamma$-ray flare to the 1995 radio outburst?
Given the superluminal motion we would expect a corresponding VLBI knot to
have a core separation of 0.3 mas after that time. In 1992.85 more than 
80\% of the 22 GHz flux was confined within 0.1 mas to the core (Pohl et
al. 1995).}
\end{figure}

\noindent
In Fig.1 we show the radio light curve of 0528+134 in comparison to its
$\gamma$-ray state for the period 1991 to 1995.
The source was quiet in the radio regime between 1985 and 1991 (Zhang et al.
1994). The following conclusion can be drawn:
0528+134 has been very bright in $\gamma$-rays either when it was weak in radio
or a few months before a mm outburst. 
It was at medium $\gamma$-ray level at the time
of the brightest ever-recorded radio outburst, at the end of 1995.
Though the $\gamma$-ray high states seem to precede the radio outbursts,
which is also supported by backextrapolation of the position of VLBI knots, 
there is no simple one-to-one relation. The analysis of individual sources
is mainly hampered by the limited coverage of the $\gamma$-ray light curve
and the problem that many sources -- like 0528+134 -- exhibit $\gamma$-ray 
variability on time scales of days, less than the standard integration
time in EGRET observations.

\section{The average source}
\noindent
We have seen that for an individual source it is difficult to relate 
$\gamma$-ray and radio outbursts to each other. This is especially true
if the source is so active that the radio outbursts blend together.

\noindent 
To circumvent this problem we may also consider the average
source.  As the absolute flux levels of sources can be different and we
are mainly interested in the variability behaviour we scale all flux
values linearly between historical minimum and historical maximum.
In this analysis we accept only sources which have been proven to be 
variable by at least a factor of two.
\begin{figure}
\plotone{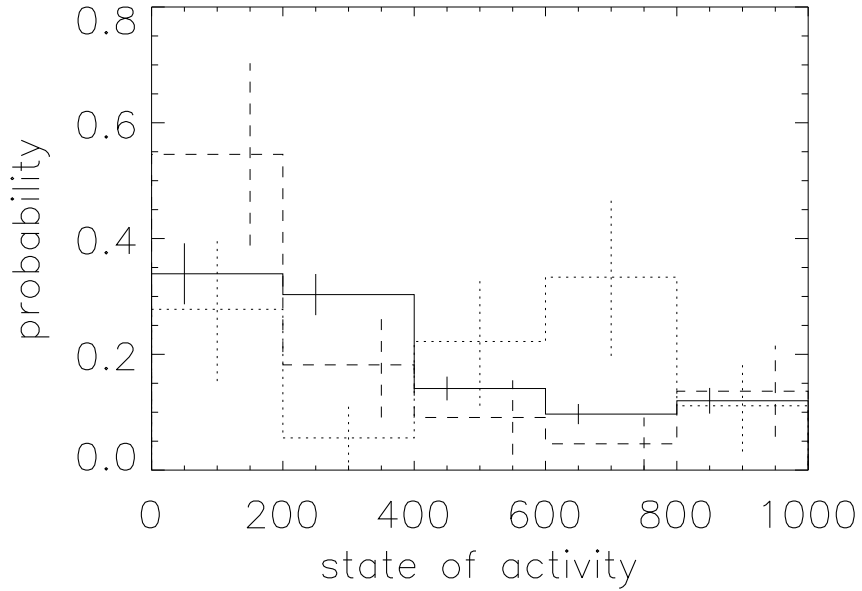}
\caption{Activity state at 10 GHz simultaneous to EGRET detections (dotted line),
averaged over the last three years for EGRET sources (solid line), and
for a sample of FSRQ which have not been seen by EGRET. Activity in
high-energy $\gamma$-rays coincides with activity in the radio regime.}
\end{figure}
\noindent
During the EGRET sky survey we have observed all blazars with cataloged
flux level $>\,$1 Jy simultaneously to the $\gamma$-ray observations.
This allows us to compare the activity distribution in the radio
regime of sources which have been seen by EGRET to those sources which
have been $\gamma$-ray quiet. Here the activity of a source is defined
on a scale of zero to thousand between historical minimum and maximum.
The resulting activity distributions are shown in Fig.2 where in case of
EGRET blazars we give both the activity at the time of $\gamma$-ray detection
and the average activity over a three years time period.
The average source has an enhanced activity level at radio wavelengths
when it is bright in high energy $\gamma$-rays (M\"ucke et al. 1996a)

\noindent
Most EGRET blazars are variable at $\gamma$-ray energies. It may be
instructive to see whether there is a relation between the radio state
and the $\gamma$-ray brightness. The $\gamma$-ray data for this analysis
are taken from the second EGRET catalog (Thompson et al. 1995) where we
have excluded observations with large off-axis angles which often yield only
poor upper limits. Then the significance of detection is used as
discriminator between $\gamma$-ray bright and $\gamma$-ray quiet.
As can be seen in Fig.3, there seems to be a clear trend,
indicating that $\gamma$-ray outbursts precede radio outbursts:
when the radio flux is increasing,
the average source is preferentially bright in $\gamma$-rays, and
when the radio flux is decreasing,
the average source is preferentially weak in $\gamma$-rays (M\"ucke et al.
1996a).
\begin{figure}
\plotone{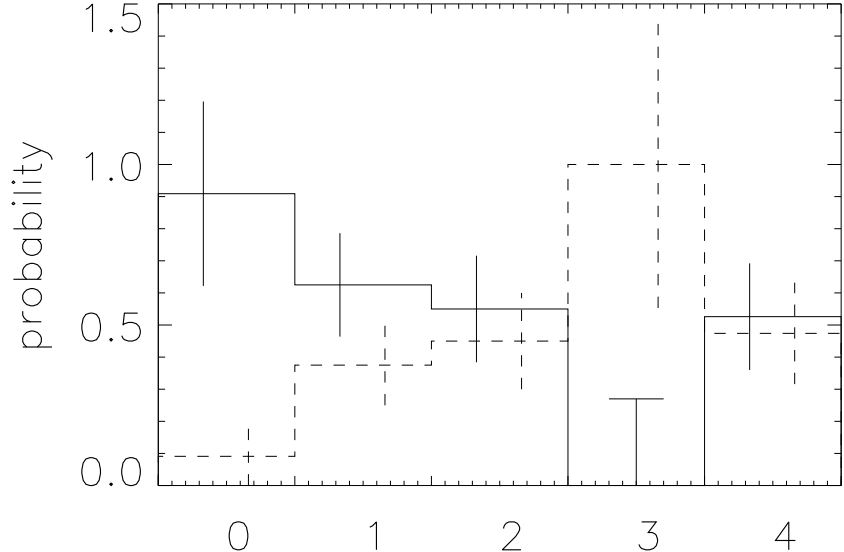}
\caption{The probability level of detection (solid line) and non-detection
(dashed line) of EGRET sources in different radio states: 0: increasing
radio flux, 1: medium level without clear trend, 2: high state, 3:
decreasing flux, 4: low state. There is a high probability level for detection
at increasing radio flux, i.e. in the rising phase of a radio outburst,
and a correspondingly small detection probability in the declining phase of
radio outbursts.}
\end{figure}
\noindent
We have also searched for correlations between radio and $\gamma$-ray 
quantities as well as between flux and spectral indices.
There is no simple relation between the scaled $\gamma$-ray flux and 
the scaled radio flux or scaled radio spectral index. 
However,
we found evidence for a relation between the $\gamma$-ray flux and
the $\gamma$-ray spectral index. This can be seen in Fig.4 where
both the $\gamma$-ray flux and the power-law spectral index have been
scaled between historical minimum and maximum.
The average source appears to
have a harder spectrum at $\gamma$-ray high states (M\"ucke et al.
1996b). The chance probability is of order $10^{-5}$ and the result is 
also stable against the omission of individual sources from the sample.

\section{Luminosity correlations}
\noindent
There is no luminosity correlation
between radio and $\gamma$-ray emission of blazars. 
\begin{figure}
\plotone{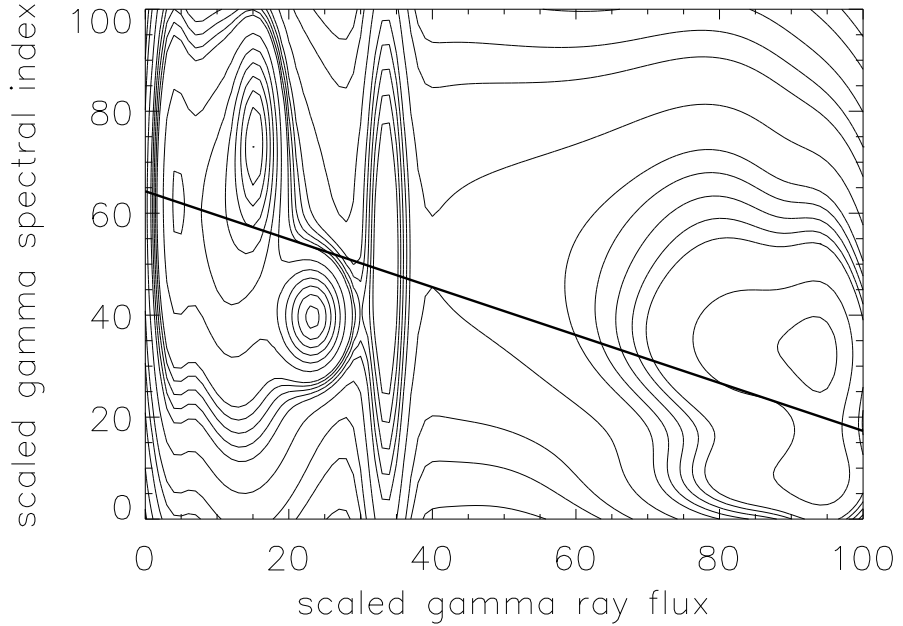}
\caption{Scaled $\gamma$-ray spectral index (0\-=\-hardest occurance
minus 1$\sigma$, 100\-=\-softest occurance plus 1$\sigma$) in
comparison to the $\gamma$-ray flux. Here the data are represented
by ellipsoidal Gaussians and the fit is based on $\chi^2$ statistic.}
\end{figure}
\noindent
We have analysed
our data with Spearman's correlation method and, to include also the
upper limits, by Kendall's $\tau$. These partial correlation coefficients
take care of the strong redshift dependence of luminosity, which may mimic
a correlation for flux-limited data. In case of the radio-to-$\gamma$-ray 
relation of blazars the apparent alignment of data in
luminosity-luminosity diagrams is indeed an artefact arising from 
sensitivity limits in the sample (M\"ucke et al. 1996c).
For EGRET observations the sensitivity
limit is mainly given by count statistic, while in the radio range mainly
a credibility limit applies, that in view of their abundance one is 
reluctant to identify a $\gamma$-ray point source with a 100 mJy
radio source, apart from the fact that catalogs including spectral information 
are not complete at these flux levels.

\noindent
The results of our analysis are summarized in Table 1.
We have also compared peak luminosities, here based on radio data
taken from the literature. The marginal correlation
signal for the 4.8 GHz data is not supported by a similar result for
the 8 GHz data and thus probably a statistical fluke.

\begin{table}
\caption{Results of the correlation analysis of N simultaneously observed
data points: partial correlation analysis using the Spearman rank order
coefficient $R_s$ or Kendall's $\tau$ for the case of censored data (i.e.
including upper limits: UL). The third column gives the resultant
correlation coefficient and the fourth column the chance probability of
erroneously rejecting the null hypothesis. An asterisk indicates that the
chance probability is determined on the basis of simulations.
The analysis was carried out
between the logarithms of the luminosities and K-corrected flux densities, 
respectively.} \label{tbl-1}
\begin{center}
\begin{tabular}{|c||c||c|c|c|}\hline
data & N & CC & prob. & correlation \\ \hline\hline
10 GHz - $\gamma$-ray (simult.) & 25 & 0.158 & 0.465 & NO \\ \hline
10 GHz - $\gamma$-ray (simult.+UL) & 42 & 0.079 & 0.288 & NO \\ \hline
2.7 GHz - $\gamma$-ray (simult.) & 22 & 0.046 & 0.751 & NO \\ \hline
2.7 GHz - $\gamma$-ray (simult.+UL) & 41 & 0.063 & 0.426 & NO \\ \hline
max. 4.8 GHz - $\gamma$-ray  & 12 & 0.594 & 0.053 & marginal \\ \hline
max. 8 GHz - $\gamma$-ray & 11 & 0.363 & 0.314 & NO \\ \hline
4.8 GHz - $\gamma$-ray obs. (mean) & 38 & 0.347 & 0.12 $\ast$& NO \\ \hline
8 GHz - $\gamma$-ray obs. (mean) & 28 & 0.405 & 0.11 $\ast$& NO \\ \hline
\end{tabular}
\end{center}
\end{table}

\noindent
When averaging over the light curves, i.e. using mean values, the
dynamical range gets even more compressed so that a positive correlation
signal is artificially produced, though the original data are uncorrelated.
Our simulations show that the chance probability for a high correlation 
coefficient is increased roughly by a factor 4 compared to the standard
$R_s$ statistic.
This may explain earlier claims for a correlation (Stecker et al. 1993;
Padovani et al. 1993).

\noindent
We also searched for linear correlations by a $\chi^2$ test on the
flux-flux relation. In no case we obtained acceptable fits of linear
relations to the observed simultaneous data.

\end{document}